\documentclass[12pt]{article}
\usepackage[latin1]{inputenc}
\usepackage[british]{babel}
\usepackage{cmap}
\usepackage{lmodern}

\usepackage{amssymb, amsmath, amsthm}
\usepackage[a4paper,top=25mm,bottom=25mm,left=25mm,right=25mm]{geometry}
\usepackage{ragged2e}

\usepackage{authblk} % for headings
\usepackage{pifont}
\usepackage{graphicx}
\usepackage[usenames,dvipsnames,svgnames,table]{xcolor}
\usepackage[figuresright]{rotating}
\usepackage{xtab} % tackle the long tables
\usepackage{longtable} % tackle the long tables
\usepackage{multirow}
\usepackage{footnote}
\usepackage[stable]{footmisc}
\usepackage{chngpage} % allows for temporary adjustment of side margins
\usepackage{pdflscape} % landscape environment
\usepackage[nottoc,notlot,notlof]{tocbibind} % includes everything in ToC

\usepackage{pgfplots}
\pgfplotsset{compat=1.14}
\pgfplotsset{every tick label/.append style={font=\footnotesize}}
\usepackage{setspace}
\makesavenoteenv{tabular}
\usepackage{tabularx}
\usepackage{booktabs}
\usepackage{threeparttable} % [flushleft]
\usepackage[referable]{threeparttablex} % footnotes in tabu
\newcolumntype{R}{>{\raggedleft\arraybackslash}X}
\newcolumntype{L}{>{\raggedright\arraybackslash}X}
\newcolumntype{C}{>{\centering\arraybackslash}X}
\newcolumntype{A}{>{\columncolor{gray!25}}C}
\newcolumntype{a}{>{\columncolor{gray!25}}c}

\newlength{\tablen}

\usepackage{dcolumn} % alignment to decimal points
\newcolumntype{.}{D{.}{.}{-1}}

\usepackage{tikz}
\usetikzlibrary{arrows, calc, matrix, patterns, positioning, trees}
\usepackage[semicolon]{natbib}
\usepackage[hyphens]{url}
\usepackage{hyperref} % [hidelinks]
\hypersetup{
  colorlinks   = true,    % Colours links instead of ugly boxes
  urlcolor     = blue,    % Colour for external hyperlinks
  linkcolor    = blue,    % Colour of internal links
  citecolor    = ForestGreen      % Colour of citations
}
\usepackage{microtype}
\usepackage[singlelinecheck=false,justification=centering]{caption}

\usepackage[labelformat=simple]{subcaption}

\DeclareCaptionLabelFormat{parenthesis}{(#2)}
\makeatletter
\renewcommand\p@subfigure{\arabic{figure}.}
\makeatother

\DeclareCaptionLabelFormat{parenthesis}{(#2)}
\makeatletter
\renewcommand\p@subtable{\arabic{table}.}
\makeatother

\usepackage{enumitem}
\setlist[itemize]{leftmargin=2.5\parindent}
\setlist[enumerate]{leftmargin=2.5\parindent}

\def\addlegendimage{\csname pgfplots@addlegendimage\endcsname}

\theoremstyle{plain}
\theoremstyle{definition}

\newtheorem{example}{Example}[section]

\theoremstyle{remark}

\def\keywords{\vspace{.5em} % Add keywords
{\noindent \textit{Keywords}: }}

\def\JEL{\vspace{.5em} % Add keywords
{\noindent \textbf{\emph{JEL} classification number}: }}

\def\AMS{\vspace{.5em} % Add keywords
{\noindent \textbf{\emph{MSC} class}: }}

\title{A comparative study of scoring systems by simulations}
\author{\href{https://sites.google.com/view/laszlocsato}{L\'aszl\'o Csat\'o}\thanks{~E-mail: \emph{laszlo.csato@sztaki.hu}} }
\affil{Institute for Computer Science and Control (SZTAKI) \\
E\"otv\"os Lor\'and Research Network (ELKH) \\
Laboratory on Engineering and Management Intelligence \\
Research Group of Operations Research and Decision Systems}
\affil{Corvinus University of Budapest (BCE) \\
Department of Operations Research and Actuarial Sciences}
\affil{Budapest, Hungary}
\date{\today}

\def\Dedication{
{\noindent
``\emph{Wer all seine Ziele erreicht, hat sie wahrscheinlich zu niedrig gew\"ahlt.}''\footnote{~
``\emph{Those who have achieved all their aims probably set them too low.}'' \\
Source: \url{http://www.karajan.org/jart/prj3/karajan/main.jart}}
}
\vspace{0.25cm}

\flushright
\noindent (Herbert von Karajan)

\vspace{1cm} 
\justify }

\begin{document}

\newgeometry{top=25mm,bottom=25mm,left=25mm,right=25mm}

\maketitle
\thispagestyle{empty}
\Dedication

\begin{abstract}
\noindent
Scoring rules aggregate individual rankings by assigning some points to each position in each ranking such that the total sum of points provides the overall ranking of the alternatives. They are widely used in sports competitions consisting of multiple contests. We study the trade-off between two risks in this setting:
(1) the threat of early clinch when the title is clinched before the last contest(s) of the competition take place; and
(2) the danger of winning the competition without finishing first in any contest.
In particular, four historical points scoring systems of the Formula One World Championship are compared with the family of geometric scoring rules, recently proposed by an axiomatic approach. The schemes used in practice are found to be competitive with respect to these goals, and the current rule seems to be a reasonable compromise close to the Pareto frontier. Our results shed more light on the evolution of the Formula One points scoring systems and contribute to the issue of choosing the set of point values.

\keywords{competition design; Formula One; OR in sports; rank aggregation; scoring system}
% sports ranking

\AMS{62F07, 68U20, 91B14}
% Ranking and selection
% Simulation
% Social choice

\JEL{C44, C63, Z20}
% Operations Research, Statistical Decision Theory
% Computational Techniques, Simulation Modeling 
% Sports Economics, General
\end{abstract}

\clearpage
\restoregeometry

\section{Introduction} \label{Sec1}

Several sporting competitions consist of a series of races. In each race, a certain number of points are assigned to the contestants based on their ranking, which are summed up across the races. Finally, the contestant with the highest score becomes the champion. This aggregation method is usually called \emph{points scoring system}. Examples include motorsport (e.g.\ \href{https://en.wikipedia.org/wiki/Formula_One}{Formula One}), road bicycle racing (e.g.\ \href{https://en.wikipedia.org/wiki/Points_classification_in_the_Tour_de_France}{Points classification in the Tour de France}), or winter sports (e.g.\ \href{https://en.wikipedia.org/wiki/Biathlon_World_Cup}{Biathlon World Cup}). Consequently, the choice of points scoring system is a fundamental question of tournament design \citep{Csato2021a}.

According to Arrow's impossibility theorem \citep{Arrow1950}, a celebrated result of social choice theory, no aggregation procedure can be considered fully satisfactory provided that the reasonable properties imposed by Arrow are accepted. For contest design, \emph{electoral consistency} seems to be a key axiom: if contestant $A$ is ranked higher or equal than contestant $B$ based on sets of races $P$ and $Q$, then $A$ should be ranked higher or equal than $B$ based on the set of races $P \cup Q$, furthermore, contestant $A$ needs to be ranked strictly higher if it is ranked strictly higher based on either $P$ or $Q$. In other words, the unique winner based on an arbitrary partition of the competition should win the whole competition. Electoral consistency can be guaranteed only by a generalised scoring rule \citep{Smith1973, Young1975}---at the price of sacrificing \emph{independence of irrelevant alternatives}, that is, the relative ranking of two contestants may depend on the performance of other contestants.

Therefore, the meticulous analysis of points scoring systems is vital not only for sports but for many other research fields like decision making, game theory, machine learning, market design, or political science.
There exist numerous axiomatic characterisations of scoring rules \citep{NitzanRubinstein1981, ChebotarevShamis1998a, Merlin2003, LlamazaresPena2015, KondratevIanovskiNesterov2022}. However, they usually result in extremities or a parametric family of rules.
Finally, it is worth noting that among all the scoring rules, the Borda rule is the least susceptible to certain
paradoxes \citep[p.~216]{BramsFishburn2002}, although it is not used in any sports competition.\footnote{~We are grateful to an anonymous referee for this remark.}

The academic literature extensively deals with the issue of aggregating individual rankings, too.
\citet{SteinMizziPfaffenberger1994} show how stochastic dominance can be applied to determine a candidate who would win using any convex scoring function and to obtain a partial ordering under a class of scoring functions.
\citet{ChurilovFlitman2006} aim to design an objective impartial system of analysing the Olympics results.
\citet{LlamazaresPena2013} propose a model to evaluate each candidate according to the most favourable weighting vector for him/her.
\citet{Sitarz2013} formulates a minimal set of conditions for points and finds the incenter of the derived convex cone.
\citet{Kaiser2019} investigates the stability of historical Formula One rankings when the parameters are slightly changed.
\citet{ZhaoWuChen2021} examine the sensitivities of weighted scoring rules via their weights and obtain some suggestions about their design.

We study the issue of assigning point values for different ranking places from a sporting perspective.
Inspired by real-world examples in Section~\ref{Sec2}, the current paper explores the trade-off between two risks:
(1) the threat of early clinch when the title has been clinched before the last contest(s) of the competition take place;
(2) the danger of winning the competition without finishing first in any contest.
Every decision-maker should find a compromise between these menaces since associating larger weights to the top ranks \emph{ceteris paribus} reduces the danger of winning without finishing first but increases the threat of early clinch.

The importance of avoiding early clinch is shown by the estimation of \citet{Garcia-del-BarrioReade2021}: around one-third of the interest in sport competitions (and the corresponding potential revenues) may be lost after the overall winner is determined. Winning at least one race can be crucial because it inspires risk-taking, which probably increases the interest of spectators and makes the competition more exciting as believed by \emph{Bernie Ecclestone}, the former head of the Formula One Group \citep{Henry2008}. In addition, lexicographic order is a standard way to rank the performance of contestants in similar competitions \citep{ChurilovFlitman2006}.

Some points scoring systems will be compared for these two criteria via simulations such that the probability of both events is quantified on the basis of Formula One World Championship results. The historical scoring systems are found to be competitive with the family of geometric scoring rules, proposed by \citet{KondratevIanovskiNesterov2022} using an axiomatic game theoretical approach.
In particular, the current points scoring system of Formula One turns out to be a good compromise for balancing the two threats.

It is worth noting that the points scoring system has serious implications for the strategy of the teams, which is also a highly relevant aspect of assigning values for different ranking positions. In order to get a tractable simulation model, these strategic decisions are neglected in our study but they have partially been discussed in the literature \citep{BekkerLotz2009, HeilmeierGrafLienkamp2018}.

Our main contribution resides in providing a useful way to choose an appropriate scoring rule for highly stochastic contests, especially where empirical evaluations based on cardinal performance are unreliable due to external factors such as wind, rain, temperature, unpredictable collisions, or mechanical breakdowns. Whereas \citet{KondratevIanovskiNesterov2022} axiomatically motivate a family of such systems, they do not study them with respect to the two unwanted scenarios considered here. Even though the data comes from one particular sport, the example may be worth attention in itself since the Formula One World Championship is widely recognised as the most prestigious motor racing competition and has a huge impact on the entire automotive industry. Hopefully, the current study will remind managers and sports governing bodies around the world that there are some inevitable trade-offs in setting the weights associated with different ranking places. We also want to foster interest in this important research question that has been addressed surprisingly rarely in social choice theory.

The paper is structured as follows.
Section~\ref{Sec2} presents two motivating examples. Section~\ref{Sec3} discusses the methodological background and underlying data. The results are provided in Section~\ref{Sec4}, while Section~\ref{Sec5} offers a concise summary and concluding remarks.

\section{Motivation} \label{Sec2}

Our starting point is an unavoidable trade-off between two objectives, illustrated by case studies from the history of sports.

\begin{table}[t]
\centering
\captionsetup{justification=centering}
\caption{The 2002 Formula One World Championship}
\label{Table1}

\begin{subtable}{\linewidth}
\centering
\caption{Points scoring system}
\label{Table1a}
%\rowcolors{1}{gray!20}{}
    \begin{tabularx}{0.6\textwidth}{l CCC CCC} \toprule
    Position & 1     & 2     & 3     & 4     & 5     & 6 \\ \midrule
    Points & 10    & 6     & 4     & 3     & 2     & 1 \\ \bottomrule
    \end{tabularx}
\end{subtable}

\vspace{0.5cm}
\begin{subtable}{\linewidth}
\caption{Race results: ranks and total scores}
\label{Table1b}
\rowcolors{1}{gray!20}{}
\centerline{
    \begin{tabularx}{1.2\textwidth}{l CCCCC CCCCC CCCCC CC >{\bfseries}c} \toprule \hiderowcolors
    \multirow{2}{*}{Driver} & \multicolumn{17}{c}{Race}                                                                                                             & Total \\
          & 1     & 2     & 3     & 4     & 5     & 6     & 7     & 8     & 9     & 10    & 11    & 12    & 13    & 14    & 15    & 16    & 17    & score \\ \bottomrule \showrowcolors
    Schumacher, M. & 1     & 3     & 1     & 1     & 1     & 1     & 2     & 1     & 2     & 1     & 1     & 1     & 2     & 1     & 2     & 2     & 1     & 144 \\
    Barrichello & ---   & ---   & ---   & 2     & ---   & 2     & 7     & 3     & 1     & 2     & ---   & 4     & 1     & 2     & 1     & 1     & 2     & 77 \\
    Montoya & 2     & 2     & 5     & 4     & 2     & 3     & ---   & ---   & ---   & 3     & 4     & 2     & 11    & 3     & ---   & 4     & 4     & 50 \\
    Schumacher, R. & ---   & 1     & 2     & 3     & 11    & 4     & 3     & 7     & 4     & 8     & 5     & 3     & 3     & 5     & ---   & 16    & 11    & 42 \\
    Coulthard & ---   & ---   & 3     & 6     & 3     & 6     & 1     & 2     & ---   & 10    & 3     & 5     & 5     & 4     & 7     & 3     & ---   & 41 \\ \toprule
    \end{tabularx}
}
\end{subtable}
\end{table}

\begin{example} \label{Examp1}
In the \href{https://en.wikipedia.org/wiki/1999_Grand_Prix_motorcycle_racing_season}{2002 Formula One World Championship}, 17 races were organised where the top six drivers scored points according to Table~\ref{Table1a}.
The outcome of the World Drivers' Championship is provided in Table~\ref{Table1b}. \emph{Michael Schumacher} clinched the title after he scored 96 points in the first 11 races, while the actual runner-up \emph{Juan-Pablo Montoya} scored only 34 points. Consequently, Schumacher had an advantage of 62 points, more than the maximal prize in the remaining six races.
\end{example}

Obviously, Example~\ref{Examp1} presents an unfavourable situation: who will be interested in watching the last race(s) if the champion is already known? This risk is called the \emph{threat of early clinch}.

\begin{table}[t]
\centering
\captionsetup{justification=centering}
\caption{The 1999 Motorcycle Grand Prix---125cc}
\label{Table2}

\begin{subtable}{\linewidth}
\caption{Points scoring system}
\label{Table2a}
%\rowcolors{1}{gray!20}{}
    \begin{tabularx}{\textwidth}{l CCCCC CCCCC CCCCC} \toprule
    Position & 1     & 2     & 3     & 4     & 5     & 6     & 7     & 8     & 9     & 10    & 11    & 12    & 13    & 14    & 15 \\ \midrule
    Points & 25    & 20    & 16    & 13    & 11    & 10    & 9     & 8     & 7     & 6     & 5     & 4     & 3     & 2     & 1 \\ \bottomrule
    \end{tabularx}
\end{subtable}

\vspace{0.5cm}
\begin{subtable}{\linewidth}
\caption{Race results: ranks and total scores}
\label{Table2b}
\rowcolors{1}{gray!20}{}
\centerline{
    \begin{tabularx}{1.1\textwidth}{l CCCCC CCCCC CCC CCC >{\bfseries}c} \toprule \hiderowcolors
    \multirow{2}{*}{Rider} & \multicolumn{16}{c}{Race}                                                                                                     & Total \\    
          & 1     & 2     & 3     & 4     & 5     & 6     & 7     & 8     & 9     & 10    & 11    & 12    & 13    & 14    & 15    & 16    & score \\ \bottomrule \showrowcolors
    Alzamora & 2     & 3     & 3     & 3     & 6     & 2     & 4     & 3     & 2     & 6     & 4     & 2     & 15    & ---   & 3     & 2     & 227 \\
    Melandri & ---   & ---   & ---   & 6     & 2     & 3     & 8     & 5     & 1     & 1     & 1     & ---   & 1     & 3     & 2     & 1     & 226 \\
    Azuma & 1     & 1     & 1     & 4     & 7     & ---   & 1     & 1     & 6     & 12    & 10    & ---   & 5     & 14    & 6     & ---   & 190 \\
    Locatelli & 18    & ---   & 5     & 1     & 1     & 6     & 3     & 4     & 4     & ---   & 11    & 8     & 6     & 4     & 8     & 3     & 173 \\
    Ueda  & ---   & ---   & 8     & 5     & 3     & 4     & 2     & 2     & 5     & 2     & 5     & 3     & ---   & ---   & 1     & ---   & 171 \\
    Scalvini & 3     & 12    & 4     & 7     & ---   & 10    & 5     & 6     & 7     & 4     & 19    & 1     & 4     & 1     & ---   & 7     & 163 \\
    Vincent & 4     & ---   & 10    & 2     & 5     & 1     & 7     & 9     & 10    & 10    & 3     & 4     & ---   & 2     & 13    & ---   & 155 \\ \toprule
    \end{tabularx}
}
\end{subtable}
\end{table}

\begin{example} \label{Examp2}
In the \href{https://en.wikipedia.org/wiki/1999_Grand_Prix_motorcycle_racing_season}{1999 Grand Prix motorcycle racing}, 16 races took place such that the top fifteen riders scored points according to Table~\ref{Table2a}.
The outcome of the 125cc category is provided in Table~\ref{Table2b}. \emph{Emilio Alzamora} was declared the champion despite not winning any race. Contrarily, both the second and the third contestants won five races.
\end{example}

Example~\ref{Examp2} outlines another situation that rule-makers probably do not like. The spectator of any sporting event wants to see a dramatic struggle for the first spot instead of risk avoidance by the runner-up. The possible situation of being the champion without winning a single race is called the \emph{danger of winning without finishing first}.

Example~\ref{Examp2} has not inspired any reform, the Motorcycle Grand Prix applies the same points scoring system since 1993.
On the other hand, Formula One has introduced a new rule from 2003 to 2009 by giving 10, 8, 6, 5, 4, 3, 2, 1 point(s) for the first eight drivers in each race, respectively. Under the latter scheme, Michael Schumacher would have scored 102 points in the first 11 races, and Montoya would have scored 50. Therefore, the runner-up still would have had some (albeit marginal) chance to grab the championship in the last six races: Michael Schumacher would have clinched the title only after 12 races.

Compared to the 2002 scoring rule, the 2003 system never provides fewer but sometimes gives more points to all drivers except for the first. Thus it has reduced the threat of early clinch---and has simultaneously enhanced the danger of winning without being first. Clearly, there exists a bargain between excitement at the level of individual races and over the whole season, which could be a potential explanation for the continuous changes in the Formula One points scoring systems \citep{Haigh2009, Kaiser2019, KondratevIanovskiNesterov2022}.

\section{Data and methodology} \label{Sec3}

\begin{table}[t!]
  \centering
  \captionsetup{justification=centering}
  \caption{The characteristics of Formula One seasons in our dataset}
  \label{Table3}
\begin{threeparttable}
\rowcolors{1}{}{gray!20}
    \begin{tabularx}{0.8\textwidth}{CCCCC} \toprule
    Season & Drivers & Races & Clinched & Margin \\ \bottomrule
    2007  & 20    & 17    & 17    & 1 \\
    2008  & 19    & 18    & 18    & 1 \\
    2009  & 20    & 17    & 16    & 11 \\
    2010  & 18    & 19    & 19    & 4 \\
    2011  & 18    & 19    & 15    & 122 \\
    2012  & 18    & 20    & 20    & 3 \\
    2013  & 18    & 19    & 16    & 155 \\
    2014  & 17    & 19    & 19    & 67 \\
    2015  & 18    & 19    & 16    & 59 \\
    2016  & 19    & 21    & 21    & 5 \\
    2017  & 19    & 20    & 18    & 46 \\
    2018  & 20    & 21    & 19    & 88 \\
    2019  & 19    & 21    & 19    & 87 \\ \toprule
    \end{tabularx}
\begin{tablenotes} \footnotesize
\item
Drivers: Number of drivers who finished in the top ten positions at the end of at least one race.
\item
Races: Number of races in the season.
\item
Clinched: Number of races after which the title of the Drivers' Championship was secured.
\item
Margin: The advantage of the world champion over the runner-up at the end of the season.
\end{tablenotes}
\end{threeparttable}
\end{table}

In order to evaluate different points scoring systems concerning the threat of early clinch and the danger of winning without finishing first, one needs hypothetical individual rankings to be aggregated. For this purpose, we start from the race results of thirteen Formula One seasons between 2007 and 2019.
Their main features are summarised in Table~\ref{Table3}.\footnote{~Some remarks on the dataset:

\href{https://en.wikipedia.org/wiki/2007_Formula_One_World_Championship}{2007}: The records of \emph{Robert Kubica} and \emph{Sebastian Vettel} were merged (the latter driver substituted the former in the seventh race); the records of \emph{Scott Speed} and \emph{Sebastian Vettel} were merged (the latter driver substituted the former in the last seven races); the records of \emph{Alexander Wurz} and \emph{Kazuki Nakajima} were merged (the latter driver substituted the former in the last race).

\href{https://en.wikipedia.org/wiki/2009_Formula_One_World_Championship}{2009}: The records of \emph{Timo Glock} and \emph{Kamui Kobayashi} were merged (the latter driver substituted the former in the last two races). \\
At the second race of the Malaysian Grand Prix, half points were awarded because less than 75\% of the scheduled distance was completed due to heavy rain.

\href{https://en.wikipedia.org/wiki/2010_Formula_One_World_Championship}{2010}: The records of \emph{Pedro de la Rosa} and \emph{Nick Heidfeld} were merged (the latter driver substituted the former in the last five races).

\href{https://en.wikipedia.org/wiki/2011_Formula_One_World_Championship}{2011}: The records of \emph{Nick Heidfeld} and \emph{Bruno Senna} were merged (the latter driver substituted the former in the last eight races).

\href{https://en.wikipedia.org/wiki/2011_Formula_One_World_Championship}{2014}: Double points were awarded in the \href{https://en.wikipedia.org/wiki/2014_Abu_Dhabi_Grand_Prix}{final race of Abu Dhabi Grand Prix}. Hence 50 points could have been collected by winning here, which explains the seemingly high margin even though the title was clinched only in the last race.

\href{https://en.wikipedia.org/wiki/2016_Formula_One_World_Championship}{2016}: The records of \emph{Fernando Alonso} and \emph{Stoffel Vandoorne} were merged (the latter driver substituted the former in the second race).

\href{https://en.wikipedia.org/wiki/2019_Formula_One_World_Championship}{2019}: One additional point was awarded for the fastest lap if the driver was classified in the top ten.
}

The drivers are ranked by the number of points scored at the end of a season. Ties are broken according to the greater number of first places, followed by the greater number of second places, and so on until a winner emerges. If the procedure fails to produce a result, the organiser nominates the winner according to such criteria as it thinks fit \citep[Article~7.2]{FIA2019}. In our simulations, only the two main ranking criteria (first: greater number of points scored; second: greater number of race wins) are implemented, with the remaining ties resolved through a drawing of lots.

The outcome of a season is described as follows.
Each driver is identified by its final standing in the World Drivers' Championship. Every race is represented by the positions of the drivers in this order. For instance, the vector $\left[ 2,\, 3,\, 1 \right]$ for a race means that the later world champion finished as the second, the later runner-up obtained the third spot, and the race was won by the driver who earned third place at the end of the season.

A race is simulated in two ways:
\begin{itemize}
\item
\emph{Method 1}: One race is drawn randomly with replacement from the set of all races included in the underlying dataset.
\item
\emph{Method 2}: Two races are drawn randomly with replacement from the set of all races included in the underlying dataset. Provisional spots are chosen by a coin toss from one of these two races, independently for each driver. This procedure is analogous to uniform crossover in genetic algorithms \citep{Syswerda1989}.
However, the provisional spots will usually not provide an appropriate ranking of the drivers. The final positions are obtained by ordering the drivers according to their provisional spots such that all ties are broken randomly.
\end{itemize}

The motivation behind Method 1 is straightforward. Method 2 can be justified by the following reasoning. An observed race represents only a single realisation of several stochastic variables, therefore using only historical results might miss some reasonable scenarios. For instance, if a driver has finished either in the first or the third position based on the data, Method 1 will never generate an outcome where this driver is the second, even though that seems to be a potential ``state of nature''.

\begin{table}[t]
\centering
\captionsetup{justification=centering}
\caption{Possible race results in Example~\ref{Examp3} under Method 2}
\label{Table4}

\begin{subtable}{\linewidth}
\centering
\caption{Provisional spots and the corresponding race results}
\label{Table4a}
\rowcolors{1}{}{gray!20}
    \begin{tabularx}{0.6\textwidth}{cCC} \toprule
    Provisional spots & \multicolumn{2}{c}{Possible race results} \\ \bottomrule
    $\left[ 1,\, 3,\, 2 \right]$ & \multicolumn{2}{c}{$\left[ 1,\, 3,\, 2 \right]$} \\
    $\left[ 1,\, 3,\, 3 \right]$ & $\left[ 1,\, 2,\, 3 \right]$ & $\left[ 1,\, 3,\, 2 \right]$ \\
    $\left[ 1,\, 1,\, 2 \right]$ & $\left[ 1,\, 2,\, 3 \right]$ & $\left[ 2,\, 1,\, 3 \right]$ \\
    $\left[ 2,\, 3,\, 2 \right]$ & $\left[ 1,\, 3,\, 2 \right]$ & $\left[ 2,\, 3,\, 1 \right]$ \\
    $\left[ 1,\, 1,\, 3 \right]$ & $\left[ 1,\, 2,\, 3 \right]$ & $\left[ 2,\, 1,\, 3 \right]$ \\
    $\left[ 2,\, 3,\, 3 \right]$ & $\left[ 1,\, 2,\, 3 \right]$ & $\left[ 1,\, 3,\, 2 \right]$ \\
    $\left[ 2,\, 1,\, 2 \right]$ & $\left[ 2,\, 1,\, 3 \right]$ & $\left[ 3,\, 1,\, 2 \right]$ \\
    $\left[ 2,\, 1,\, 3 \right]$ & \multicolumn{2}{c}{$\left[ 2,\, 1,\, 3 \right]$} \\ \bottomrule
    \end{tabularx}
\end{subtable}

\vspace{0.5cm}
\begin{subtable}{\linewidth}
\centering
\caption{Race results and their probabilities}
\label{Table4b}
\rowcolors{1}{}{gray!20}
    \begin{tabularx}{0.8\textwidth}{cC} \toprule
    Race result & Probability \\ \bottomrule
    $\left[ 1,\, 2,\, 3 \right]$ & $1/16+1/16+1/16+1/16=4/16=1/4$ \\
    $\left[ 1,\, 3,\, 2 \right]$ & $1/8+1/16+1/16+1/16=5/16$ \\
    $\left[ 2,\, 1,\, 3 \right]$ & $1/16+1/16+1/16+1/8=5/16$ \\
    $\left[ 2,\, 3,\, 1 \right]$ & $1/16$ \\
    $\left[ 3,\, 1,\, 2 \right]$ & $1/16$ \\
    $\left[ 3,\, 2,\, 1 \right]$ & $0$ \\ \bottomrule
    \end{tabularx}
\end{subtable}
\end{table}

\begin{example} \label{Examp3}
Assume that there are three drivers and two races in the underlying dataset, $A = \left[ 1,\, 3,\, 2 \right]$ and $B = \left[ 2,\, 1,\, 3 \right]$. Consequently, the champion has one first and one second place, the runner-up has one first and one third place, while the remaining contestant has one second and one third place in the races, which is possible under any monotonic scoring rule that does not give fewer points for a better finishing spot.
The result of a race simulated according to Method 1 can be either $A$ or $B$.
In the case of Method 2, there are eight ($2^3$) possible lists of provisional spots because the number of drivers is three, each of them occurring with a probability of $1/8$. Table~\ref{Table4a} details them together with the corresponding race results. The probabilities of feasible race results are reported in Table~\ref{Table4b}.

Note that the second driver has finished as second neither in race $A$ nor in race $B$ but this event has a 25\% chance to happen under Method 2.
\end{example}

%Note that in Examplean observed race represents only a single realisation of several stochastic variables. Therefore, Method 2, which is able to generate a more abundant set of scenarios compared to Method 1, might better represent the potential ``states of nature''.

A season that contains a given number of races is attained by generating the appropriate number of individual race results independently from each other.

\begin{table}[t]
  \centering
  \caption{The points scoring systems of the analysis}
  \label{Table5}
\rowcolors{1}{gray!20}{}
\centerline{
    \begin{tabularx}{1.05\textwidth}{c CCCCC CCCCC} \toprule \hiderowcolors
    \multirow{2}{*}{Rule} & \multicolumn{10}{c}{Position} \\
         & 1     & 2     & 3     & 4     & 5     & 6     & 7     & 8     & 9     & 10 \\ \bottomrule \showrowcolors
    $S1$ & 9     & 6     & 4     & 3     & 2     & 1     & 0     & 0     & 0     & 0 \\
    $S2$ & 10    & 6     & 4     & 3     & 2     & 1     & 0     & 0     & 0     & 0 \\
    $S3$ & 10    & 8     & 6     & 5     & 4     & 3     & 2     & 1     & 0     & 0 \\
    $S4$ & 25    & 18    & 15    & 12    & 10    & 8     & 6     & 4     & 2     & 1 \\
    $G1$ & 10    & 9     & 8     & 7     & 6     & 5     & 4     & 3     & 2     & 1 \\
    $G2$ & 12.58 & 11.03 & 9.55  & 8.14  & 6.80  & 5.53  & 4.310 & 3.1525 & 2.05  & 1 \\
    $G3$ & 42.62 & 32.01 & 23.86 & 17.58 & 12.76 & 9.04  & 6.187 & 3.99  & 2.3   & 1 \\
    $G4$ & 181.59 & 112.87 & 69.92 & 43.07 & 26.30 & 15.81 & 9.256 & 5.16  & 2.6   & 1 \\ \bottomrule
    \end{tabularx}
}
\end{table}

Eight points scoring rules, shown in Table~\ref{Table5}, are considered:
\begin{itemize}
\item
\emph{System $S1$}: the official Formula One points scoring system between 1961 and 1990.
Although not all results counted in the drivers' championship \citep{Kaiser2019}, this restriction is not implemented in our simulations.
\item
\emph{System $S2$}: the official points scoring system of Formula One between 1991 and 2002.
\item
\emph{System $S3$}: the official points scoring system of Formula One between 2003 and 2009.
\item
\emph{System $S4$}: the official Formula One points scoring system since 2010.
Two complications were the double points awarded in the \href{https://en.wikipedia.org/wiki/2014_Abu_Dhabi_Grand_Prix}{last race of the 2014 season} and one additional point for the fastest lap in 2019 if the driver was classified in the top ten. However, they do not appear in our simulations.
\item
\emph{Systems $G1$--$G4$}: according to \citet[Theorem~2]{KondratevIanovskiNesterov2022}, consistency for adding or removing a unanimous winner or loser pins down the one-parameter family \emph{geometric scoring rules} with the set of scores $1$, $1+p$, $1+p+p^2$, \dots \\
Again, points are awarded to the top ten spots, thus the score $s_j$ of the $j$th position is $\left( p^{11-j} - 1 \right) / (p-1)$ for all $1 \leq j \leq 10$ if $p > 1$, which---because the scores are not distinguished up to scaling and translation---is equivalent to $\hat{s_j} = p^{11-j}$. Four different values of the parameter have been chosen. $G1$ is a relatively ``flat'', linearly increasing scoring scheme with $p=1$. $G2$ denotes the case $p = 1.05$, $G3$ denotes the case $p = 1.3$, and $G4$ denotes the case $p = 1.6$.
\end{itemize}

\begin{figure}[t]

\begin{tikzpicture}
\begin{axis}[
name = axis1,
xlabel = Positions,
x label style = {font=\small},
width = \textwidth,
height = 0.6\textwidth,
ymajorgrids = true,
xmin = 1,
xmax = 10,
ymin = 0,
ymax = 100,
max space between ticks=50,
legend style = {font=\small,at={(0.05,-0.15)},anchor=north west,legend columns=8},
legend entries = {$S1 \quad$,$S2 \quad$,$S3 \quad$,$S4 \quad$,$G1 \quad$,$G2 \quad$,$G3 \quad$,$G4$}
] 

% S1
\addplot [brown, thick, mark=oplus*] coordinates {
(1,100)
(2,66.6666666666667)
(3,44.4444444444444)
(4,33.3333333333333)
(5,22.2222222222222)
(6,11.1111111111111)
(7,0)
(8,0)
(9,0)
(10,0)
};

% S2
\addplot [gray, thick, mark=diamond*, mark options={thick}] coordinates {
(1,100)
(2,60)
(3,40)
(4,30)
(5,20)
(6,10)
(7,0)
(8,0)
(9,0)
(10,0)
};

% S3
\addplot [orange, thick, mark=triangle*, mark options={solid,thick}] coordinates {
(1,100)
(2,80)
(3,60)
(4,50)
(5,40)
(6,30)
(7,20)
(8,10)
(9,0)
(10,0)
};

% S4
\addplot [ForestGreen, thick, mark=square*, mark options={thin}] coordinates {
(1,100)
(2,72)
(3,60)
(4,48)
(5,40)
(6,32)
(7,24)
(8,16)
(9,8)
(10,4)
};

% F
\addplot [red, thick, mark=pentagon*] coordinates {
(1,100)
(2,90)
(3,80)
(4,70)
(5,60)
(6,50)
(7,40)
(8,30)
(9,20)
(10,10)
};

% G1
\addplot [blue, thick, dashdotted, mark=asterisk, mark options={solid,thick}] coordinates {
(1,100)
(2,87.6662309556708)
(3,75.9197842467858)
(4,64.7326921430859)
(5,54.0783187109906)
(6,43.9312963947095)
(7,34.2674656172989)
(8,25.0638172578602)
(9,16.2984378679186)
(10,7.95045749654567)
};

% G2
\addplot [black, thick, densely dotted, mark=x, mark options={solid,very thick}] coordinates {
(1,100)
(2,75.1181969543552)
(3,55.9783484577053)
(4,41.255388075667)
(5,29.9300339356375)
(6,21.2182230586917)
(7,14.5168300764258)
(8,9.36191239775962)
(9,5.39659110647797)
(10,2.34634395933825)
};

% G3
\addplot [purple, thick, dashed, mark=+, mark options={solid,very thick}] coordinates {
(1,100)
(2,62.1558090887332)
(3,38.5031897691914)
(4,23.7203026944778)
(5,14.4809982727818)
(6,8.70643300922185)
(7,5.09732971949685)
(8,2.84164016341873)
(9,1.4318341908699)
(10,0.550705458026885)
};
\end{axis}
\end{tikzpicture}

\captionsetup{justification=centerfirst}
\caption{The points scoring systems of Table~\ref{Table5} \\ \vspace{0.25cm}
\footnotesize{Scores for first place are normalised to 100.}}
\label{Fig1}

\end{figure}
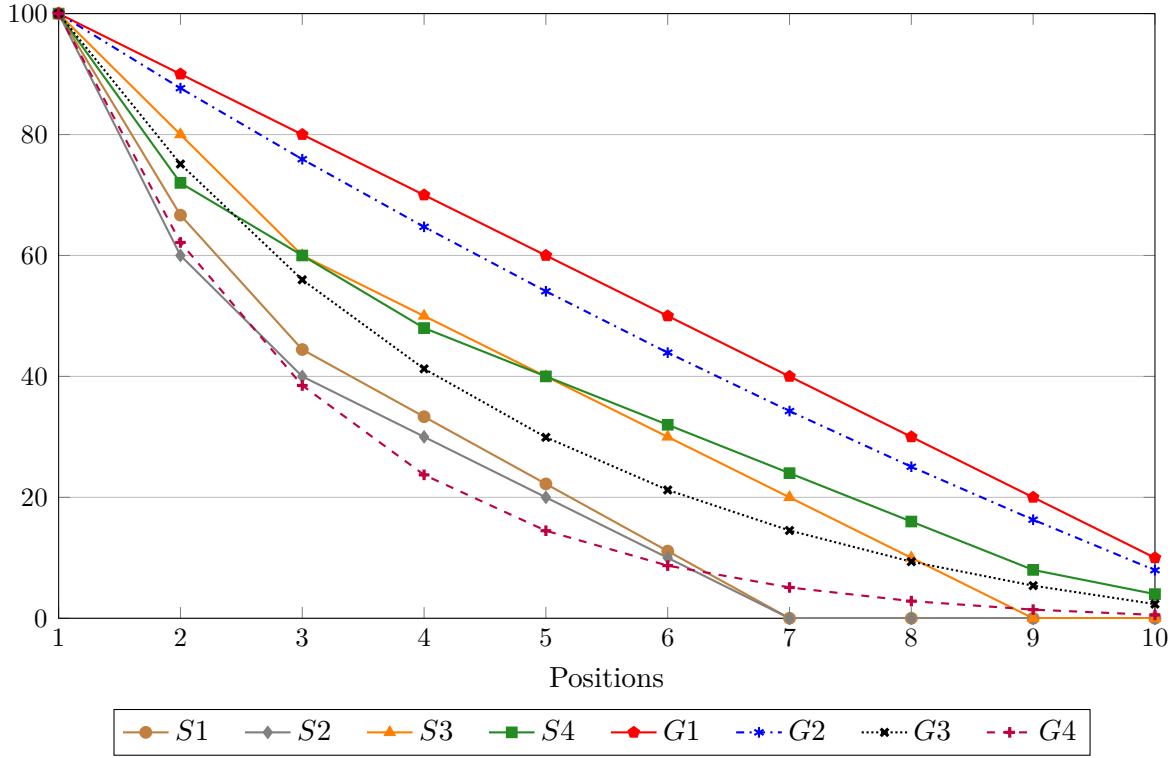

%\end{document}

The eight points scoring systems are plotted in Figure~\ref{Fig1} such that the scores for the first spot are scaled to 100.

Two sets of years are studied:
\begin{itemize}
\item
\emph{Standard dataset}: all races of every season between 2010 and 2019.
\item
\emph{Small margin dataset}: all races of the seasons 2007, 2008, 2009, 2010, 2012, and 2016.
\end{itemize}
The standard dataset contains ten recent seasons, where the rules were almost the same (scoring system $S4$). Hence, it can be reasonably assumed that the incentives of the drivers did not change substantially during these years.
As the column entitled ``Margin'' in Table~\ref{Table3} uncovers, the small margin dataset consists of the seasons when the difference between the first two drivers was marginal at the end, thus the competition was balanced and open until the last race(s). Note that in 2014, the championship was open in the last race despite the seemingly high margin of $67$ points as the winner got $50$ points there and \emph{Lewis Hamilton} led by $17$ points over \emph{Nico Rosberg} before this race.

According to preliminary simulations, winning without finishing first in any race rarely occurs if a season of reasonable length is generated on the basis of the above data. It is probably caused by the steep points scoring systems used between 2007 and 2019 ($S3$ and $S4$) that provide powerful incentives for taking risks in order to win a race. However, this motivation is not so strong if a relatively flat scoring rule is applied to mitigate the threat of early clinch. In addition, even a completely risk averse contestant may win a race out of many merely by chance as the following example shows.\footnote{~We are grateful to an anonymous referee for providing us with this excellent real-world illustration.}

\begin{table}[t]
\centering
\captionsetup{justification=centering}
\caption{The 2020 MotoGP World Championship: ranks and total scores}
\label{Table6}
\begin{threeparttable}
\rowcolors{1}{gray!20}{}
\centerline{
    \begin{tabularx}{1\textwidth}{l CCCCC CCCCC CCCC >{\bfseries}c} \toprule \hiderowcolors
    \multirow{2}{*}{Rider} & \multicolumn{14}{c}{Race}                                                                                                     & Total \\    
          & 1     & 2     & 3     & 4     & 5     & 6     & 7     & 8     & 9     & 10    & 11    & 12    & 13    & 14    & score \\ \bottomrule \showrowcolors
    Mir   & ---   & 5     & ---   & 2     & 4     & 3     & 2     & 2     & 11    & 3     & 3     & 1     & 7     & ---   & 171 \\
    Morbidelli & 5     & ---   & 2     & ---   & 15    & 1     & 9     & 4     & ---   & 6     & 1     & 11    & 1     & 3     & 158 \\
    Rins  & ---   & 10    & 4     & ---   & 6     & 5     & 12    & 3     & ---   & 1     & 2     & 2     & 4     & 15    & 139 \\
    Dovizioso & 3     & 6     & 11    & 1     & 5     & 7     & 8     & ---   & 4     & 7     & 13    & 8     & 8     & 6     & 135 \\
    Espargar\'o & 6     & 7     & ---   & ---   & 3     & 10    & 3     & ---   & 3     & 12    & 4     & 3     & 3     & 4     & 135 \\
    Vi{\~ n}ales & 2     & 2     & 14    & 10    & ---   & 6     & 1     & 9     & 10    & 4     & 7     & 13    & 10    & 11    & 132 \\
    Miller & 4     & ---   & 9     & 3     & 2     & 8     & ---   & 5     & ---   & 9     & ---   & 6     & 2     & 2     & 132 \\ \toprule
    \end{tabularx}
}
\begin{tablenotes} \footnotesize
\item
See Table~\ref{Table2a} for the points scoring system used in the season.
\end{tablenotes}
\end{threeparttable}
\end{table}

\begin{example} \label{Examp4}
In the \href{https://en.wikipedia.org/wiki/2020_MotoGP_World_Championship}{2020 MotoGP World Championship}, there were 14 races and the top fifteen riders scored points according to Table~\ref{Table2a}.
The final standing is given in Table~\ref{Table6}. \emph{Joan Mir} secured the title at the penultimate round with 29 points ahead of \emph{Franco Morbidelli}. Although he won the 12th race, he would have remained the champion even if he would have finished fourth in that particular race.
\end{example}

Consequently, the real data does not necessarily reflect the threat of winning without finishing first in an appropriate way. Hence an alternative set of season results is generated: after a whole season is simulated, all first spots of the world champion---who is determined by the current points scoring system $S4$---are turned into a second spot and the corresponding runner-up of each affected race becomes the winner.

\begin{example} \label{Examp5}
Assume that a season consists of two races, $A = \left[ 1,\, 3,\, 2 \right]$ and $B = \left[ 2,\, 1,\, 3 \right]$. The first driver obtains the championship due to having a first and a second position.
The risk averse season result is $\left[ 2,\, 3,\, 1 \right]$ and $\left[ 2,\, 1,\, 3 \right]$, namely, the winner and the runner-up in race $A$ are reversed but the outcome of race $B$ does not change because the first driver is only the runner-up there. The first driver could not remain the champion without finishing first in any race under any points scoring system of Table~\ref{Table5} since the prize for a first and a third position is never less than the prize for two second spots, and the tie-breaking rule is the number of races won.
\end{example}
The algorithm above aims to maximise the probability of winning without finishing first in a greedy way by assuming that the original world champion does not take any risk to win a race.

A simulation run contains the following phases:
\begin{enumerate}
\item
Choice of the underlying dataset: standard or small margin.
\item
Choice of the race generation procedure: method 1 or method 2.
\item \label{Step3}
Calculating the alternative season result on the basis of the original season result: generating $n$ races independently, and reversing the race winner and the runner-up in each race if the former is the champion under scoring rule $S4$.
\item \label{Step4}
Identification of the champion under the eight scoring rules listed in Table~\ref{Table5}.
\item \label{Step5}
Checking whether the champion has won at least one race.
\item \label{Step6}
Determining the last race when the title has not already been secured, that is, checking whether the difference between the scores of the champion and the actual runner-up (who is not necessarily the final runner-up) after $m$ races is smaller than the product of $n-m$ and the score awarded to the first spot, or they are equal but the difference between the number of race wins for the champion and for the actual runner-up is at most $n-m$. One of these conditions holds if and only if the title is still not clinched after $m$ races.
\end{enumerate}
The process is carried out 100 thousand times independently in each simulation.

Naturally, our approach to simulate individual rankings has several limitations. However, it should be kept in mind that---in contrast to some statistical studies \citep{GravesReeseFitzgerald2003, HendersonKirrane2018}---we do not aim to predict Formula One results and estimate the chance of a driver to win. For the evaluation of different points scoring systems, essentially any reasonable model can be taken to determine the rankings \citep{Appleton1995, Csato2021b}. Nonetheless, the qualitative findings would be more reliable than the exact numerical values, and the calculations below are mainly for comparative purposes.
%Nonetheless, this implies that the following calculations are mainly for comparative purposes.

\section{Results} \label{Sec4}

\begin{figure}[t!]

\begin{tikzpicture}
%\selectcolormodel{gray}
\begin{axis}[width = 1\textwidth, 
height = 0.65\textwidth,
ymajorgrids,
xlabel = Average number of uninteresting races,
xlabel style = {font = \small},
ylabel style={align=center},
ylabel = {Probability that the champion did not win any race},
ylabel style = {font = \small},
legend entries = {Method 1: Standard dataset$\qquad$,Method 1: Small margin dataset,Method 2: Standard dataset$\qquad$,Method 2: Small margin dataset},
legend style = {at = {(0.5,-0.15)},anchor = north,legend columns = 2,font = \small}
]

% Method 1, standard seasons
\addplot[red,thick,only marks,mark=diamond*, mark size=3pt] coordinates {
(1.23083,0.14297)
(1.61869,0.06172)
(0.67628,0.35853)
(0.95785,0.14971)
(0.52018,0.56536)
(0.53705,0.52924)
(0.8184,0.27786)
(1.44843,0.10604)
};

% Method 1, small margin
\addplot[blue,thick,only marks,mark=star, mark size=3pt] coordinates {
(0.68067,0.14391)
(0.84035,0.06639)
(0.42301,0.36457)
(0.49354,0.17306)
(0.43276,0.55843)
(0.41777,0.53589)
(0.47868,0.27689)
(0.78756,0.09929)
};

% Method 2, standard seasons
\addplot[black,thick,only marks,mark=otimes*, mark size=2pt] coordinates {
(1.18112,0.17088)
(1.54424,0.07919)
(0.68854,0.393)
(0.9251,0.1775)
(0.55653,0.58568)
(0.57102,0.55392)
(0.81078,0.31036)
(1.38234,0.13109)
};

% Method 2, small margin
\addplot[ForestGreen,very thick,only marks,mark=x, mark size=3pt] coordinates {
(0.70779,0.15682)
(0.87612,0.07504)
(0.44944,0.38131)
(0.50795,0.19135)
(0.45997,0.57415)
(0.4479,0.55298)
(0.49958,0.29643)
(0.8202,0.11113)
};

% Nodes: Method 2, standard seasons
\draw (1.18112,0.17088) node [right] {\footnotesize{$S1$}};
\draw (1.54424,0.07919) node [right] {\footnotesize{$S2$}};
\draw (0.68854,0.393)   node [left]  {\footnotesize{$S3$}};
\draw (0.9251,0.1775)   node [right] {\footnotesize{$S4$}};
\draw (0.55653,0.58568) node [left]  {\footnotesize{$G1$}};
\draw (0.57102,0.55392) node [below] {\footnotesize{$G2$}};
\draw (0.81078,0.31036) node [right] {\footnotesize{$G3$}};
\draw (1.38234,0.13109) node [right] {\footnotesize{$G4$}};

% Nodes: Method 2, small margin
\draw (0.70779,0.15682) node [above left]  {\footnotesize{$S1$}}; 
\draw (0.87612,0.07504) node [below] {\footnotesize{$S2$}}; % Method 1
\draw (0.44944,0.38131) node [above left]  {\footnotesize{$S3$}}; 
\draw (0.50795,0.19135) node [left]  {\footnotesize{$S4$}};
\draw (0.43276,0.55843) node [above] {\footnotesize{$G1$}}; % Method 1
\draw (0.41777,0.53589) node [right] {\footnotesize{$G2$}}; % Method 1
\draw (0.49958,0.29643) node [left]  {\footnotesize{$G3$}};
\draw (0.8202,0.11113)  node [above left] {\footnotesize{$G4$}}; % Method 1
\end{axis}
\end{tikzpicture}

\captionsetup{justification=centering}
\caption{The tradeoff between the threat of early clinch and the \\ danger of winning without finishing first (20 races per season)}
\label{Fig2}

\end{figure}
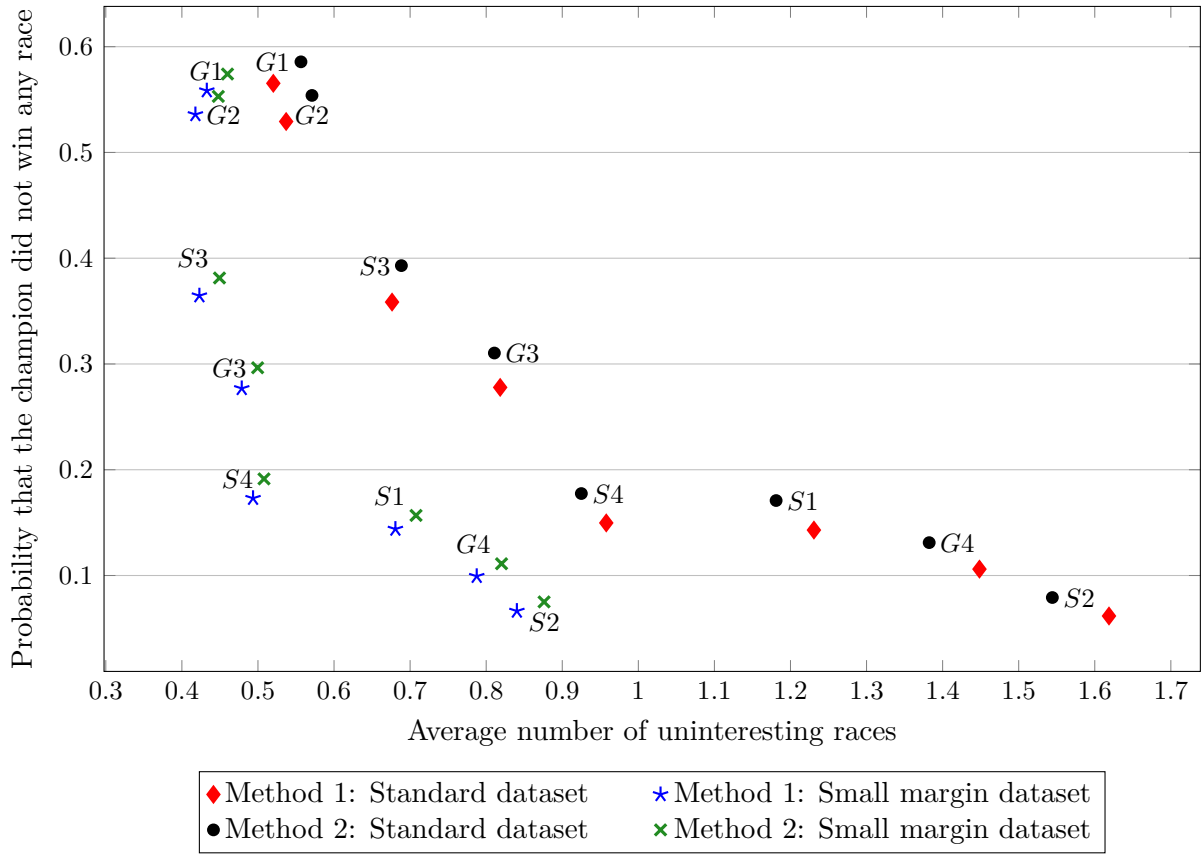

%\end{document}

A race can be called \emph{uninteresting} if the title has already been clinched by a contestant.
Figure~\Ref{Fig2} plots the probability that the champion did not win any race during a season consisting of 20 races as a function of the average number of these uninteresting races. Since there is essentially no difference between methods 1 and 2, only the latter will be considered in the following.
%While the choice of the dataset marginally influences the danger of winning without finishing first, it has an important role concerning the threat of early clinch as expected.

According to Figure~\ref{Fig2}, the two goals can be optimised only at the expense of each other.
While schemes $S1$ and $S2$ are very similar at first sight because only the number of points awarded to the first spot is modified (see Table~\ref{Table5}), this has a significant effect on both of our measures. For example, $S1$ doubles the danger of winning without finishing first compared to $S2$.
Even though geometric scoring rules have strong theoretical foundations and are not vulnerable to removing unanimous winners or losers \citep{KondratevIanovskiNesterov2022}, they do not outperform the historical points scoring systems.

It is worth evoking the ``evolution'' of the official points scoring system from $S1$ to $S4$. The change from $S2$ to $S3$ was probably inspired by Example~\ref{Examp1} because the title was clinched by Michael Schumacher after 11 races out of the total 17 in 2002 (and after 13 races out of 17 in the previous year). Understandably, the organisers chose option $S3$ to decrease the threat of early clinch. However, it was a kind of overreaction by remarkably reducing the role of winning. The designers finally reached the compromise $S4$, which is about halfway between $S1$ and $S3$ from the aspect of winning without finishing first. There are further interesting findings in Figure~\ref{Fig2}:
\begin{itemize}
\item
Scoring rule $S4$ outperforms $S3$ on the basis of small margin seasons (2007--2010, 2012, 2016), and $S3$ was changed to $S4$, effective from 2010. Perhaps the administrators recognised that the danger of winning without finishing first can be substantially decreased at a small sacrifice in the threat of early clinch.
\item
Compared to $S4$, system $S1$ yields only a small reduction in the probability that the champion did not win any race, while the average number of uninteresting races significantly increases. This can explain why $S1$ did not return as the official rule in the 2000s.
\item
The average number of uninteresting races has perhaps a natural minimum as this measure cannot be effectively decreased by making the scores flatter.
\item
System $S4$ is probably closer to the Pareto frontier than all other schemes, especially on the basis of the small margin dataset, that is, in the most competitive seasons.
\end{itemize}
To conclude, the current points scoring rule $S4$ seems to perform well with respect to both objectives.

\begin{figure}[t!]

\begin{tikzpicture}
% Left axis
\begin{axis}[width=0.98\textwidth, 
height=0.5\textwidth,
title = Standard dataset,
title style = {font=\small},
tick label style={/pgf/number format/fixed},
symbolic x coords={$S1$,$S2$,$S3$,$S4$,$G1$,$G2$,$G3$,$G4$},
xtick = data,
%xlabel = Points scoring system,
%xlabel style = {font = \small},
enlarge x limits={abs=1cm},
axis y line* = left,
ybar,
ymin = 0,
ymax = 1.6,
bar width = 0.5cm,
ybar = 0.1cm,
max space between ticks = 40,
]

\addplot [blue, pattern color = blue, pattern = dots, very thick] coordinates{
($S1$,1.18112)
($S2$,1.54424)
($S3$,0.68854)
($S4$,0.9251)
($G1$,0.55653)
($G2$,0.57102)
($G3$,0.81078)
($G4$,1.38234)
};

\addplot [red, pattern color = red, pattern = vertical lines, very thick] coordinates{
($S1$,0)
($S2$,0)
($S3$,0)
($S4$,0)
($G1$,0)
($G2$,0)
($G3$,0)
($G4$,0)
};
\end{axis}

% Right axis
\begin{axis}[width=0.98\textwidth, 
height=0.5\textwidth,
%title = Full dataset,
%tick label style={/pgf/number format/fixed},
symbolic x coords={$S1$,$S2$,$S3$,$S4$,$G1$,$G2$,$G3$,$G4$},
xticklabels={,,},
%xlabel = Points scoring system,
%xlabel style = {font = \small},
enlarge x limits={abs=1cm},
axis y line* = right,
ybar,
ymin = 0,
ymax = 0.32,
ymajorgrids = true,
yticklabel style={/pgf/number format/fixed, /pgf/number format/precision=2},
%ymajorgrids = true,
bar width = 0.5cm,
ybar = 0.1cm,
]

\addplot [blue, pattern color = blue, pattern = dots, very thick] coordinates{
($S1$,0)
($S2$,0)
($S3$,0)
($S4$,0)
($G1$,0)
($G2$,0)
($G3$,0)
($G4$,0)
};

\addplot [red, pattern color = red, pattern = vertical lines, very thick] coordinates{
($S1$,0.13069)
($S2$,0.23394)
($S3$,0.02392)
($S4$,0.06217)
($G1$,0.01276)
($G2$,0.01408)
($G3$,0.04302)
($G4$,0.1889)
};
\end{axis}
\end{tikzpicture}

\vspace{0.5cm}
\begin{tikzpicture}
% Left axis
\begin{axis}[width=0.98\textwidth, 
height=0.5\textwidth,
title = Small margin dataset,
title style = {font=\small},
tick label style={/pgf/number format/fixed},
symbolic x coords={$S1$,$S2$,$S3$,$S4$,$G1$,$G2$,$G3$,$G4$},
xtick = data,
%xlabel = Points scoring system,
%xlabel style = {font = \small},
enlarge x limits={abs=1cm},
axis y line* = left,
ybar,
ymin = 0,
ymax = 1.05,
ytick distance = 0.25,
%max space between ticks = 40,
%ymajorgrids = true,
bar width = 0.5cm,
ybar = 0.1cm,
legend entries = {Average number of uninteresting races (left scale)$\qquad \qquad \qquad \quad$,Probability that at least three races are uninteresting (right scale)},
legend style = {at={(0.5,-0.15)},anchor = north,legend columns = 1,font=\small}
]

\addplot [blue, pattern color = blue, pattern = dots, very thick] coordinates{
($S1$,0.70779)
($S2$,0.87612)
($S3$,0.44944)
($S4$,0.50795)
($G1$,0.45997)
($G2$,0.4479)
($G3$,0.49958)
($G4$,0.8202)
};

\addplot [red, pattern color = red, pattern = vertical lines, very thick] coordinates{
($S1$,0)
($S2$,0)
($S3$,0)
($S4$,0)
($G1$,0)
($G2$,0)
($G3$,0)
($G4$,0)
};
\end{axis}

% Right axis
\begin{axis}[width=0.98\textwidth, 
height=0.5\textwidth,
%title = Full dataset,
%tick label style={/pgf/number format/fixed},
symbolic x coords={$S1$,$S2$,$S3$,$S4$,$G1$,$G2$,$G3$,$G4$},
xticklabels={,,},
%xlabel = Points scoring system,
%xlabel style = {font = \small},
enlarge x limits={abs=1cm},
axis y line* = right,
scaled ticks = false,
y tick label style = {/pgf/number format/.cd,fixed,precision=3},
ybar,
ymin = 0,
ymax = 0.084,
%max space between ticks = 50,
ymajorgrids = true,
bar width = 0.5cm,
ybar = 0.1cm,
]

\addplot [blue, pattern color = blue, pattern = dots, very thick] coordinates{
($S1$,0)
($S2$,0)
($S3$,0)
($S4$,0)
($G1$,0)
($G2$,0)
($G3$,0)
($G4$,0)
};

\addplot [red, pattern color = red, pattern = vertical lines, very thick] coordinates{
($S1$,0.03892)
($S2$,0.06963)
($S3$,0.00943)
($S4$,0.01232)
($G1$,0.01173)
($G2$,0.01138)
($G3$,0.01251)
($G4$,0.05929)
};
\end{axis}
\end{tikzpicture}

\captionsetup{justification=centering}
\caption{The threat of early clinch (20 races per season, method 2)}
\label{Fig3}

\end{figure}
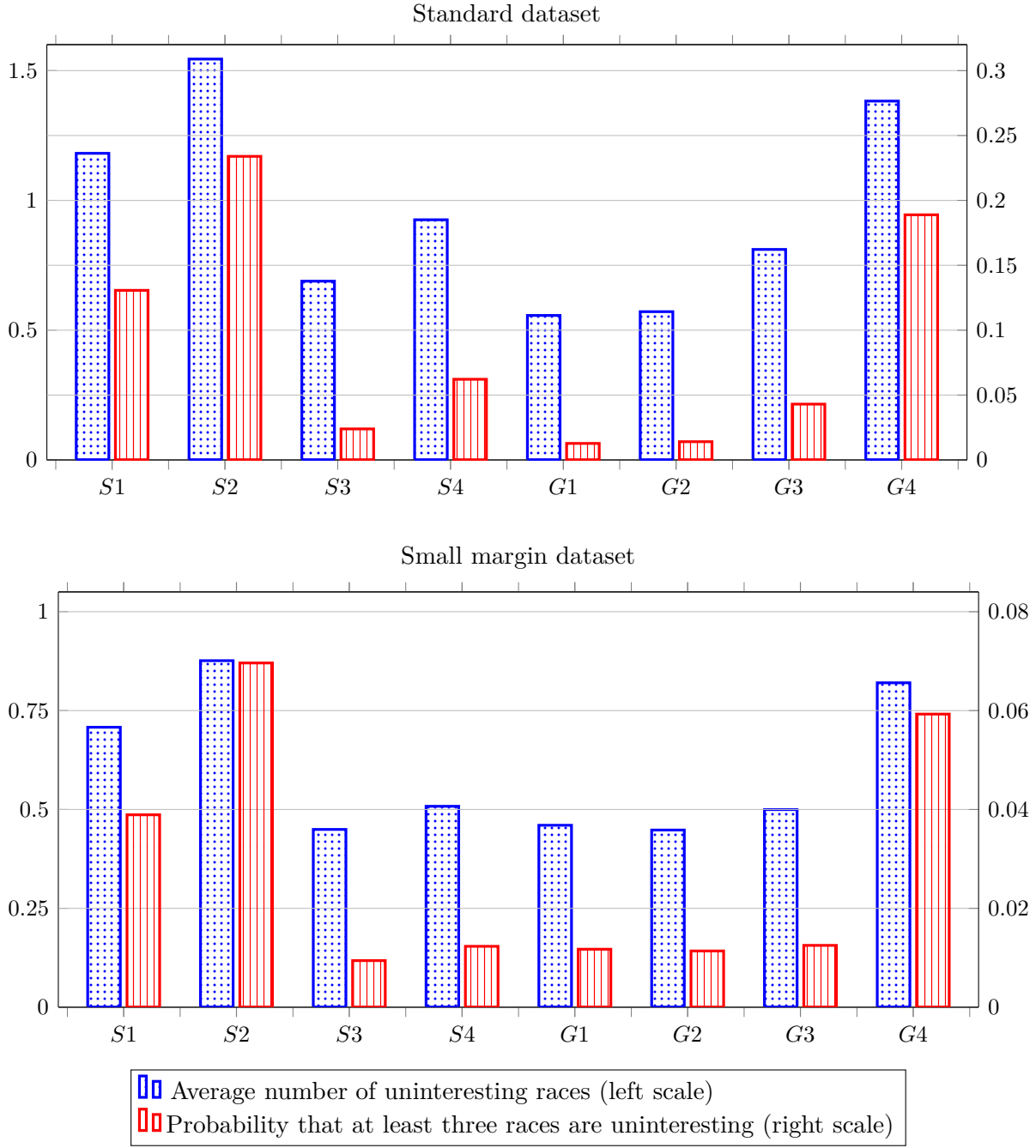

%\end{document}

Figure~\ref{Fig3} quantifies the threat of early clinch by the average number of races when the title has already been secured and the probability that the champion is decided before the last three races out of the total 20. It reinforces the previous findings, the geometric scoring rule $G4$ is similar to $S1$ and $S2$, while $G3$ might be a good alternative to $S3$ or $S4$ from the viewpoint of early clinches. The threat of early clinch cannot be eliminated, schemes $G1$ is at most marginally better than $G2$ according to both measures---naturally, by facilitating the risk that the title is grabbed without finishing first in any race.

\begin{figure}[t!]

\begin{subfigure}{\textwidth}
\captionsetup{justification=centering}
\caption{Average number of uninteresting races}
\label{Fig4a}

\begin{tikzpicture}
\begin{axis}[
name = axis1,
title = {Standard dataset},
title style = {font=\small},
xlabel = Number of races per season,
x label style = {font=\small},
width = 0.5\textwidth,
height = 0.4\textwidth,
ymajorgrids = true,
xmin = 4,
xmax = 20,
ymin = 0,
%ymax = 100,
%max space between ticks=50,
%legend style = {font=\small,at={(0.05,-0.15)},anchor=north west,legend columns=8},
%legend entries = {$S1 \quad$,$S2 \quad$,$S3 \quad$,$S4 \quad$,$F \quad$,$G1 \quad$,$G2 \quad$,$G3$}
] 

% S1
\addplot [ForestGreen, thick, mark=square*, mark options={thin}] coordinates {
(3,0.02868)
(4,0.09113)
(5,0.12075)
(6,0.17638)
(7,0.23171)
(8,0.29855)
(9,0.3647)
(10,0.43334)
(11,0.50461)
(12,0.57895)
(13,0.65086)
(14,0.72844)
(15,0.80309)
(16,0.87486)
(17,0.94339)
(18,1.02591)
(19,1.09534)
(20,1.18112)
};

% S2
\addplot [gray, thick, mark=diamond*, mark options={thick}] coordinates {
(3,0.07417)
(4,0.10416)
(5,0.18153)
(6,0.23945)
(7,0.32079)
(8,0.39902)
(9,0.48395)
(10,0.57279)
(11,0.66179)
(12,0.75804)
(13,0.84819)
(14,0.94725)
(15,1.04761)
(16,1.14458)
(17,1.23281)
(18,1.33845)
(19,1.4369)
(20,1.54424)
};

% S3
\addplot [brown, thick, mark=oplus*] coordinates {
(3,0.00072)
(4,0.01294)
(5,0.04216)
(6,0.07455)
(7,0.11264)
(8,0.1559)
(9,0.19915)
(10,0.24094)
(11,0.2866)
(12,0.33148)
(13,0.3754)
(14,0.41984)
(15,0.46738)
(16,0.50937)
(17,0.55117)
(18,0.5994)
(19,0.64584)
(20,0.68854)
};

% S4
\addplot [orange, thick, mark=triangle*, mark options={solid,thick}] coordinates {
(3,0.00072)
(4,0.01935)
(5,0.06194)
(6,0.10288)
(7,0.14534)
(8,0.19598)
(9,0.24676)
(10,0.30078)
(11,0.35919)
(12,0.42071)
(13,0.47874)
(14,0.54243)
(15,0.60703)
(16,0.66588)
(17,0.72491)
(18,0.7928)
(19,0.85289)
(20,0.9251)
};

% G1
\addplot [red, thick, mark=pentagon*] coordinates {
(3,0.00001)
(4,0.00143)
(5,0.00996)
(6,0.03207)
(7,0.06249)
(8,0.10032)
(9,0.13981)
(10,0.17693)
(11,0.21685)
(12,0.25542)
(13,0.29389)
(14,0.33151)
(15,0.36941)
(16,0.40839)
(17,0.44537)
(18,0.48318)
(19,0.52336)
(20,0.55653)
};

% G2
\addplot [blue, thick, dashdotted, mark=asterisk, mark options={solid,thick}] coordinates {
(3,0.00006)
(4,0.00211)
(5,0.01349)
(6,0.04061)
(7,0.07521)
(8,0.11353)
(9,0.15327)
(10,0.19103)
(11,0.23132)
(12,0.26952)
(13,0.30727)
(14,0.34641)
(15,0.38433)
(16,0.42089)
(17,0.46004)
(18,0.49801)
(19,0.53863)
(20,0.57102)
};

% G3
\addplot [black, thick, densely dotted, mark=x, mark options={solid,very thick}] coordinates {
(3,0.00302)
(4,0.02259)
(5,0.05487)
(6,0.09194)
(7,0.13737)
(8,0.18766)
(9,0.23618)
(10,0.28323)
(11,0.33691)
(12,0.38984)
(13,0.44225)
(14,0.49367)
(15,0.54953)
(16,0.59842)
(17,0.64692)
(18,0.70618)
(19,0.75659)
(20,0.81078)
};

% G4
\addplot [purple, thick, dashed, mark=+, mark options={solid,very thick}] coordinates {
(3,0.02868)
(4,0.09322)
(5,0.13434)
(6,0.20128)
(7,0.27243)
(8,0.35105)
(9,0.42849)
(10,0.50878)
(11,0.59242)
(12,0.67869)
(13,0.76018)
(14,0.84986)
(15,0.94149)
(16,1.02357)
(17,1.10433)
(18,1.20117)
(19,1.28618)
(20,1.38234)
};
%\legend{}
\end{axis}

\begin{axis}[
at = {(axis1.south east)},
xshift = 0.1\textwidth,
title = {Small margin dataset},
title style = {font=\small},
xlabel = Number of races per season,
x label style = {font=\small},
width = 0.5\textwidth,
height = 0.4\textwidth,
ymajorgrids = true,
xmin = 4,
xmax = 20,
ymin = 0,
]

% S1
\addplot [ForestGreen, thick, mark=square*, mark options={thin}] coordinates {
(3,0.0319)
(4,0.04675)
(5,0.07549)
(6,0.10532)
(7,0.14017)
(8,0.17713)
(9,0.21631)
(10,0.2563)
(11,0.29476)
(12,0.33633)
(13,0.37633)
(14,0.42273)
(15,0.469)
(16,0.5132)
(17,0.56276)
(18,0.61098)
(19,0.66277)
(20,0.70779)
};

% S2
\addplot [gray, thick, mark=diamond*, mark options={thick}] coordinates {
(3,0.06484)
(4,0.05642)
(5,0.10013)
(6,0.13656)
(7,0.17744)
(8,0.22181)
(9,0.26952)
(10,0.31648)
(11,0.36367)
(12,0.41734)
(13,0.46835)
(14,0.52432)
(15,0.58206)
(16,0.63597)
(17,0.69737)
(18,0.75717)
(19,0.82074)
(20,0.87612)
};

% S3
\addplot [brown, thick, mark=oplus*] coordinates {
(3,0.00278)
(4,0.0144)
(5,0.0369)
(6,0.06424)
(7,0.09245)
(8,0.11857)
(9,0.14609)
(10,0.17561)
(11,0.19961)
(12,0.2289)
(13,0.25367)
(14,0.28226)
(15,0.31248)
(16,0.33897)
(17,0.36801)
(18,0.39321)
(19,0.42331)
(20,0.44944)
};

% S4
\addplot [orange, thick, mark=triangle*, mark options={solid,thick}] coordinates {
(3,0.00278)
(4,0.0163)
(5,0.0328)
(6,0.05563)
(7,0.08109)
(8,0.10512)
(9,0.13397)
(10,0.1658)
(11,0.19403)
(12,0.2278)
(13,0.25518)
(14,0.2913)
(15,0.32814)
(16,0.36198)
(17,0.39851)
(18,0.43166)
(19,0.47458)
(20,0.50795)
};

% G1
\addplot [red, thick, mark=pentagon*] coordinates {
(3,0.00005)
(4,0.00278)
(5,0.01921)
(6,0.05087)
(7,0.08446)
(8,0.1157)
(9,0.14636)
(10,0.17987)
(11,0.208)
(12,0.23997)
(13,0.26735)
(14,0.29761)
(15,0.32732)
(16,0.35084)
(17,0.38477)
(18,0.40445)
(19,0.43486)
(20,0.45997)
};

% G2
\addplot [blue, thick, dashdotted, mark=asterisk, mark options={solid,thick}] coordinates {
(3,0.00014)
(4,0.00459)
(5,0.02594)
(6,0.0592)
(7,0.09134)
(8,0.12201)
(9,0.15121)
(10,0.18297)
(11,0.20983)
(12,0.23956)
(13,0.26458)
(14,0.29334)
(15,0.32217)
(16,0.34483)
(17,0.37599)
(18,0.39535)
(19,0.42281)
(20,0.4479)
};

% G3
\addplot [black, thick, densely dotted, mark=x, mark options={solid,very thick}] coordinates {
(3,0.00758)
(4,0.02112)
(5,0.0437)
(6,0.07157)
(7,0.09955)
(8,0.12626)
(9,0.15508)
(10,0.18684)
(11,0.21353)
(12,0.24347)
(13,0.27172)
(14,0.30509)
(15,0.33833)
(16,0.37061)
(17,0.40372)
(18,0.43235)
(19,0.47002)
(20,0.49958)
};

% G4
\addplot [purple, thick, dashed, mark=+, mark options={solid,very thick}] coordinates {
(3,0.0319)
(4,0.04694)
(5,0.08733)
(6,0.11991)
(7,0.16077)
(8,0.20453)
(9,0.25011)
(10,0.29527)
(11,0.33949)
(12,0.3893)
(13,0.43864)
(14,0.49043)
(15,0.54463)
(16,0.59542)
(17,0.65449)
(18,0.70903)
(19,0.76747)
(20,0.8202)
};
\end{axis}
\end{tikzpicture}
\end{subfigure}

\vspace{0.5cm}
\begin{subfigure}{\textwidth}
\captionsetup{justification=centering}
\caption{Probability that the champion did not win any race}
\label{Fig4b}

\begin{tikzpicture}
\begin{axis}[
name = axis2,
title = {Standard dataset},
title style = {font=\small},
xlabel = Number of races per season,
x label style = {font=\small},
width = 0.5\textwidth,
height = 0.4\textwidth,
ymajorgrids = true,
xmin = 4,
xmax = 20,
ymin = 0,
ymax = 0.65,
%max space between ticks=50,
%legend style = {font=\small,at={(0.05,-0.15)},anchor=north west,legend columns=8},
%legend entries = {$S1 \quad$,$S2 \quad$,$S3 \quad$,$S4 \quad$,$F \quad$,$G1 \quad$,$G2 \quad$,$G3$}
] 

% S1
\addplot [ForestGreen, thick, mark=square*, mark options={thin}] coordinates {
(3,0.22183)
(4,0.22959)
(5,0.23251)
(6,0.22974)
(7,0.22396)
(8,0.21759)
(9,0.21702)
(10,0.21079)
(11,0.20616)
(12,0.20205)
(13,0.19939)
(14,0.19295)
(15,0.18729)
(16,0.18509)
(17,0.18125)
(18,0.17822)
(19,0.17417)
(20,0.17088)
};

% S2
\addplot [gray, thick, mark=diamond*, mark options={thick}] coordinates {
(3,0.15753)
(4,0.1287)
(5,0.13995)
(6,0.13324)
(7,0.13012)
(8,0.12351)
(9,0.12103)
(10,0.11594)
(11,0.11192)
(12,0.10792)
(13,0.1046)
(14,0.09968)
(15,0.09523)
(16,0.09174)
(17,0.08972)
(18,0.08582)
(19,0.08197)
(20,0.07919)
};

% S3
\addplot [brown, thick, mark=oplus*] coordinates {
(3,0.41232)
(4,0.42333)
(5,0.42383)
(6,0.42664)
(7,0.42422)
(8,0.41818)
(9,0.41845)
(10,0.41622)
(11,0.41357)
(12,0.41007)
(13,0.4102)
(14,0.40622)
(15,0.4015)
(16,0.40212)
(17,0.402)
(18,0.39805)
(19,0.39806)
(20,0.393)
};

% S4
\addplot [orange, thick, mark=triangle*, mark options={solid,thick}] coordinates {
(3,0.28818)
(4,0.25873)
(5,0.25293)
(6,0.24576)
(7,0.2382)
(8,0.23011)
(9,0.22763)
(10,0.22197)
(11,0.21541)
(12,0.20873)
(13,0.20665)
(14,0.20027)
(15,0.1931)
(16,0.19181)
(17,0.18781)
(18,0.18405)
(19,0.18114)
(20,0.1775)
};

% G1
\addplot [red, thick, mark=pentagon*] coordinates {
(3,0.51265)
(4,0.53691)
(5,0.55368)
(6,0.56045)
(7,0.56271)
(8,0.56341)
(9,0.56963)
(10,0.57212)
(11,0.57391)
(12,0.57391)
(13,0.57892)
(14,0.5786)
(15,0.57847)
(16,0.58349)
(17,0.585)
(18,0.58385)
(19,0.5861)
(20,0.58568)
};

% G2
\addplot [blue, thick, dashdotted, mark=asterisk, mark options={solid,thick}] coordinates {
(3,0.51264)
(4,0.53943)
(5,0.55512)
(6,0.55977)
(7,0.56049)
(8,0.55954)
(9,0.56373)
(10,0.56301)
(11,0.56202)
(12,0.55899)
(13,0.56172)
(14,0.55866)
(15,0.55441)
(16,0.55854)
(17,0.55738)
(18,0.55468)
(19,0.55624)
(20,0.55392)
};

% G3
\addplot [black, thick, densely dotted, mark=x, mark options={solid,very thick}] coordinates {
(3,0.39891)
(4,0.40067)
(5,0.39077)
(6,0.37704)
(7,0.37174)
(8,0.36096)
(9,0.35811)
(10,0.35275)
(11,0.3469)
(12,0.34034)
(13,0.3393)
(14,0.3324)
(15,0.32639)
(16,0.32441)
(17,0.32259)
(18,0.31745)
(19,0.31614)
(20,0.31036)
};

% G4
\addplot [purple, thick, dashed, mark=+, mark options={solid,very thick}] coordinates {
(3,0.12686)
(3,0.23046)
(4,0.23068)
(5,0.21635)
(6,0.207)
(7,0.19849)
(8,0.18941)
(9,0.18517)
(10,0.17756)
(11,0.17257)
(12,0.16723)
(13,0.16298)
(14,0.15671)
(15,0.15005)
(16,0.14626)
(17,0.14268)
(18,0.13905)
(19,0.13417)
(20,0.13109)
};
%\legend{}
\end{axis}

\begin{axis}[
at = {(axis2.south east)},
xshift = 0.1\textwidth,
title = {Small margin dataset},
title style = {font=\small},
xlabel = Number of races per season,
x label style = {font=\small},
width = 0.5\textwidth,
height = 0.4\textwidth,
ymajorgrids = true,
xmin = 4,
xmax = 20,
ymin = 0,
ymax = 0.65,
%max space between ticks=50,
legend style = {font=\small,at={(-1.1,-0.25)},anchor=north west,legend columns=8},
legend entries = {$S1 \quad$,$S2 \quad$,$S3 \quad$,$S4 \quad$,$G1 \quad$,$G2 \quad$,$G3 \quad$,$G4$}
]

% S1
\addplot [ForestGreen, thick, mark=square*, mark options={thin}] coordinates {
(3,0.28406)
(4,0.27482)
(5,0.27129)
(6,0.25878)
(7,0.24818)
(8,0.23862)
(9,0.23093)
(10,0.22226)
(11,0.2146)
(12,0.20645)
(13,0.19846)
(14,0.19095)
(15,0.18526)
(16,0.18029)
(17,0.17254)
(18,0.1682)
(19,0.16087)
(20,0.15682)
};

% S2
\addplot [gray, thick, mark=diamond*, mark options={thick}] coordinates {
(3,0.20715)
(4,0.17429)
(5,0.16913)
(6,0.16033)
(7,0.15136)
(8,0.14209)
(9,0.13511)
(10,0.12732)
(11,0.12082)
(12,0.1139)
(13,0.10622)
(14,0.10081)
(15,0.09775)
(16,0.09153)
(17,0.08749)
(18,0.08288)
(19,0.07851)
(20,0.07504)
};

% S3
\addplot [brown, thick, mark=oplus*] coordinates {
(3,0.49461)
(4,0.48728)
(5,0.48478)
(6,0.47314)
(7,0.4656)
(8,0.46017)
(9,0.45097)
(10,0.44412)
(11,0.43637)
(12,0.43068)
(13,0.42216)
(14,0.41676)
(15,0.41065)
(16,0.40441)
(17,0.39744)
(18,0.39104)
(19,0.38652)
(20,0.38131)
};

% S4
\addplot [orange, thick, mark=triangle*, mark options={solid,thick}] coordinates {
(3,0.38044)
(4,0.35281)
(5,0.33444)
(6,0.3155)
(7,0.30146)
(8,0.28845)
(9,0.27762)
(10,0.26697)
(11,0.25623)
(12,0.25149)
(13,0.23925)
(14,0.23094)
(15,0.22324)
(16,0.21698)
(17,0.20891)
(18,0.20208)
(19,0.19628)
(20,0.19135)
};

% G1
\addplot [red, thick, mark=pentagon*] coordinates {
(3,0.62873)
(4,0.62168)
(5,0.61791)
(6,0.60776)
(7,0.60394)
(8,0.60298)
(9,0.59888)
(10,0.59459)
(11,0.59213)
(12,0.59123)
(13,0.58921)
(14,0.58618)
(15,0.58463)
(16,0.5841)
(17,0.58046)
(18,0.57834)
(19,0.57742)
(20,0.57415)
};

% G2
\addplot [blue, thick, dashdotted, mark=asterisk, mark options={solid,thick}] coordinates {
(3,0.62853)
(4,0.62649)
(5,0.62565)
(6,0.61493)
(7,0.60868)
(8,0.6072)
(9,0.59917)
(10,0.59266)
(11,0.58971)
(12,0.58523)
(13,0.58194)
(14,0.5766)
(15,0.57267)
(16,0.57012)
(17,0.56394)
(18,0.56065)
(19,0.55845)
(20,0.55298)
};

% G3
\addplot [black, thick, densely dotted, mark=x, mark options={solid,very thick}] coordinates {
(3,0.46973)
(4,0.45466)
(5,0.44311)
(6,0.42369)
(7,0.41004)
(8,0.39848)
(9,0.387)
(10,0.3755)
(11,0.36536)
(12,0.35913)
(13,0.34612)
(14,0.33834)
(15,0.33091)
(16,0.32307)
(17,0.31474)
(18,0.3077)
(19,0.30075)
(20,0.29643)
};

% G4
\addplot [purple, thick, dashed, mark=+, mark options={solid,very thick}] coordinates {
(3,0.29225)
(4,0.25687)
(5,0.24271)
(6,0.22468)
(7,0.21107)
(8,0.19763)
(9,0.18815)
(10,0.17888)
(11,0.17001)
(12,0.16108)
(13,0.15182)
(14,0.14449)
(15,0.13997)
(16,0.13293)
(17,0.12694)
(18,0.12209)
(19,0.11571)
(20,0.11113)
};
\end{axis}
\end{tikzpicture}
\end{subfigure}

\captionsetup{justification=centering}
\caption{The threat of early clinch and the danger of winning without finishing first \\ as the function of the number of races per season (method 2)}
\label{Fig4}

\end{figure}

%\end{document}

Finally, the connection between the two dangers and the number of races is uncovered in Figure~\ref{Fig4}. The average number of uninteresting races grows almost linearly with the length of the season. In addition, Figure~\ref{Fig4} reinforces our conjecture based on Figure~\ref{Fig2} that there exists a natural ``lower bound'' for this measure. On the other hand, the danger of winning without finishing first decreases when there are more races, with the possible exception of the ``flat'' scoring rules $G1$ and $G2$. The relative performance of geometric scoring schemes $G3$ and $G4$ improves as the number of races grows but the current rule $S4$ pursues them well.
However, since even a completely risk averse contestant may win a race out of many merely by chance, our risk averse scenario may be farther from reality if the season contains more races.

\section{Conclusions} \label{Sec5}

This research has attempted to help to understand the properties of scoring rules, extensively applied to derive an aggregate ranking from individual rankings, from a sporting perspective. In particular, we have focused on the trade-off between two risks in a championship composed of multiple contests: the threat of early clinch (when the title is secured before all contests are over, therefore the last contest(s) become uninteresting) and the danger of winning without finishing first (when the overall winner does not win any contest). The probability of both events has been measured using Formula One World Championship results. The historical points scoring systems are found to be competitive with the family of geometric scoring rules, recommended by \citet{KondratevIanovskiNesterov2022} based on an axiomatic game theoretical approach. In particular, the current points scoring system of Formula One is shown to be a good compromise for balancing the two perils.

There are straightforward directions for future research.
The robustness of the simulations can be improved by considering other datasets. The danger of winning without finishing first can be quantified in alternative ways. Other aspects of adopting a points scoring system can be built into the framework. For instance, our focus has been limited to winning without finishing first, while geometric rules with a high parameter $p$ encourage risk taking at \emph{any} position.

In some competitions with a similar format, the ranking is decided by totalling the time the contestants take on each race with possible bonuses and penalties for winning races and infractions of the rules, see e.g.\ \href{https://en.wikipedia.org/wiki/General_classification_in_the_Tour_de_France}{General classification in the Tour de France}. Then the two risks addressed above are hard to interpret. However, there will likely be debates on the ``value'' of the races, the tradeoff between a one-minute lead in one race and a one-minute lead in another race. For example, the importance of Tour de France stages varies greatly \citep{Wilcockson2022}. Therefore, a simulation methodology can be used to determine the decisiveness of the races, which is somewhat analogous to measuring the role of individual games in a tournament consisting of several matches \citep{Geenens2014, CoronaHorrilloWiper2017}.

Hopefully, this work will turn out to be only the first step in the comparison of scoring rules via simulations and will serve to foster interest in a significant research question that has been addressed surprisingly rarely.
 
\section*{Acknowledgements}
\addcontentsline{toc}{section}{Acknowledgements}
\noindent
This paper could not have been written without \emph{my father} (also called \emph{L\'aszl\'o Csat\'o}), who has coded the simulations in Python. \\
We are grateful to \emph{Aleksei Y.~Kondratev} and \emph{Josep Freixas} for inspiration. \\
\emph{Aleksei Y. Kondratev}, \emph{D\'ora Gr\'eta Petr\'oczy}, and \emph{Mike Yearworth} provided valuable comments and suggestions on earlier drafts. \\
\emph{D\'ora Gr\'eta Petr\'oczy} and \emph{Gerg{\H o} T\'ov\'ari} helped in data collection. \\
%\emph{Alex Krumer}, \emph{M\'anuel M\'ag\'o} and \emph{Tam\'as Halm} gave useful comments. \\
Nine anonymous reviewer gave useful remarks on earlier drafts. \\
We are indebted to the \href{https://en.wikipedia.org/wiki/Wikipedia_community}{Wikipedia community} for contributing to our research by summarising important details of the sports competitions discussed in the paper. \\
The research was supported by the MTA Premium Postdoctoral Research Program grant PPD2019-9/2019.

\bibliographystyle{apalike}
\bibliography{All_references}

\begin{thebibliography}{}

\bibitem[Appleton, 1995]{Appleton1995}
Appleton, D.~R. (1995).
\newblock May the best man win?
\newblock {\em Journal of the Royal Statistical Society: Series D (The
  Statistician)}, 44(4):529--538.

\bibitem[Arrow, 1950]{Arrow1950}
Arrow, K.~J. (1950).
\newblock A difficulty in the concept of social welfare.
\newblock {\em Journal of Political Economy}, 58(4):328--346.

\bibitem[Bekker and Lotz, 2009]{BekkerLotz2009}
Bekker, J. and Lotz, W. (2009).
\newblock Planning {F}ormula {O}ne race strategies using discrete-event
  simulation.
\newblock {\em Journal of the Operational Research Society}, 60(7):952--961.

\bibitem[Brams and Fishburn, 2002]{BramsFishburn2002}
Brams, S.~J. and Fishburn, P.~C. (2002).
\newblock Voting procedures.
\newblock In Arrow, K.~J., Sen, A.~K., and Suzumura, K., editors, {\em Handbook
  of Social Choice and Welfare}, volume~1, chapter~4, pages 173--236. Elsevier,
  Amsterdam, The Netherlands.

\bibitem[Chebotarev and Shamis, 1998]{ChebotarevShamis1998a}
Chebotarev, P.~{\relax Yu}. and Shamis, E. (1998).
\newblock Characterizations of scoring methods for preference aggregation.
\newblock {\em Annals of Operations Research}, 80:299--332.

\bibitem[Churilov and Flitman, 2006]{ChurilovFlitman2006}
Churilov, L. and Flitman, A. (2006).
\newblock Towards fair ranking of {O}lympics achievements: The case of {S}ydney
  2000.
\newblock {\em Computers \& Operations Research}, 33(7):2057--2082.

\bibitem[Corona et~al., 2017]{CoronaHorrilloWiper2017}
Corona, F., Tena, J.~d.~D., and Wiper, M.~P. (2017).
\newblock On the importance of the probabilistic model in identifying the most
  decisive games in a tournament.
\newblock {\em Journal of Quantitative Analysis in Sports}, 13(1):11--23.

\bibitem[Csat\'o, 2021a]{Csato2021b}
Csat\'o, L. (2021a).
\newblock A simulation comparison of tournament designs for the {W}orld {M}en's
  {H}andball {C}hampionships.
\newblock {\em International Transactions in Operational Research},
  28(5):2377--2401.

\bibitem[Csat\'o, 2021b]{Csato2021a}
Csat\'o, L. (2021b).
\newblock {\em Tournament Design: How Operations Research Can Improve Sports
  Rules}.
\newblock Palgrave Pivots in Sports Economics. Palgrave Macmillan, Cham,
  Switzerland.

\bibitem[FIA, 2019]{FIA2019}
FIA (2019).
\newblock {\em 2019 Formula One Sporting Regulations}.
\newblock 12 March.
  \url{https://www.fia.com/sites/default/files/2019_sporting_regulations_-_2019-03-12.pdf}.

\bibitem[Garcia-del Barrio and Reade, 2021]{Garcia-del-BarrioReade2021}
Garcia-del Barrio, P. and Reade, J.~J. (2021).
\newblock Does certainty on the winner diminish the interest in sport
  competitions? {T}he case of formula one.
\newblock {\em Empirical Economics}, in press.
\newblock {DOI}:
  \href{https://doi.org/10.1007/s00181-021-02147-8}{10.1007/s00181-021-02147-8}.

\bibitem[Geenens, 2014]{Geenens2014}
Geenens, G. (2014).
\newblock On the decisiveness of a game in a tournament.
\newblock {\em European Journal of Operational Research}, 232(1):156--168.

\bibitem[Graves et~al., 2003]{GravesReeseFitzgerald2003}
Graves, T., Reese, C.~S., and Fitzgerald, M. (2003).
\newblock Hierarchical models for permutations: Analysis of auto racing
  results.
\newblock {\em Journal of the American Statistical Association},
  98(462):282--291.

\bibitem[Haigh, 2009]{Haigh2009}
Haigh, J. (2009).
\newblock Uses and limitations of mathematics in sport.
\newblock {\em IMA Journal of Management Mathematics}, 20(2):97--108.

\bibitem[Heilmeier et~al., 2018]{HeilmeierGrafLienkamp2018}
Heilmeier, A., Graf, M., and Lienkamp, M. (2018).
\newblock A race simulation for strategy decisions in circuit motorsports.
\newblock In {\em 21st International Conference on Intelligent Transportation
  Systems (ITSC)}, pages 2986--2993.

\bibitem[Henderson and Kirrane, 2018]{HendersonKirrane2018}
Henderson, D.~A. and Kirrane, L.~J. (2018).
\newblock A comparison of truncated and time-weighted {P}lackett--{L}uce models
  for probabilistic forecasting of {F}ormula {O}ne results.
\newblock {\em Bayesian Analysis}, 13(2):335--358.

\bibitem[Henry, 2008]{Henry2008}
Henry, A. (2008).
\newblock Ecclestone insists new {F1} medal system will be adopted by next
  season.
\newblock {\em The Guardian}.
\newblock 26 November.
  \url{https://www.theguardian.com/sport/2008/nov/26/bernie-ecclestone-formula-one-world-championship-points-medals}.

\bibitem[Kaiser, 2019]{Kaiser2019}
Kaiser, B. (2019).
\newblock Strategy and paradoxes of {B}orda count in {F}ormula 1 racing.
\newblock {\em Decyzje}, 6(31):115--132.

\bibitem[Kondratev et~al., 2022]{KondratevIanovskiNesterov2022}
Kondratev, A.~Y., Ianovski, E., and Nesterov, A.~S. (2022).
\newblock How should we score athletes and candidates: geometric scoring rules.
\newblock Manuscript. {DOI}:
  \href{https://doi.org/10.48550/arXiv.1907.05082}{10.48550/arXiv.1907.05082}.

\bibitem[Llamazares and Pe{\~n}a, 2013]{LlamazaresPena2013}
Llamazares, B. and Pe{\~n}a, T. (2013).
\newblock Aggregating preferences rankings with variable weights.
\newblock {\em European Journal of Operational Research}, 230(2):348--355.

\bibitem[Llamazares and Pe{\~n}a, 2015]{LlamazaresPena2015}
Llamazares, B. and Pe{\~n}a, T. (2015).
\newblock Scoring rules and social choice properties: some characterizations.
\newblock {\em Theory and Decision}, 78(3):429--450.

\bibitem[Merlin, 2003]{Merlin2003}
Merlin, V. (2003).
\newblock The axiomatic characterizations of majority voting and scoring rules.
\newblock {\em Math{\'e}matiques et Sciences Humaines}, 41(161):87--109.

\bibitem[Nitzan and Rubinstein, 1981]{NitzanRubinstein1981}
Nitzan, S. and Rubinstein, A. (1981).
\newblock A further characterization of {B}orda ranking method.
\newblock {\em Public Choice}, 36(1):153--158.

\bibitem[Sitarz, 2013]{Sitarz2013}
Sitarz, S. (2013).
\newblock The medal points' incenter for rankings in sport.
\newblock {\em Applied Mathematics Letters}, 26(4):408--412.

\bibitem[Smith, 1973]{Smith1973}
Smith, J.~H. (1973).
\newblock Aggregation of preferences with variable electorate.
\newblock {\em Econometrica}, 26(1):1027--1041.

\bibitem[Stein et~al., 1994]{SteinMizziPfaffenberger1994}
Stein, W.~E., Mizzi, P.~J., and Pfaffenberger, R.~C. (1994).
\newblock A stochastic dominance analysis of ranked voting systems with
  scoring.
\newblock {\em European Journal of Operational Research}, 74(1):78--85.

\bibitem[Syswerda, 1989]{Syswerda1989}
Syswerda, G. (1989).
\newblock Uniform crossover in genetic algorithms.
\newblock In Schaffer, J.~D., editor, {\em Proceedings of the 3rd International
  Conference on Genetic Algorithms}, pages 2--9.

\bibitem[Wilcockson, 2022]{Wilcockson2022}
Wilcockson, J. (2022).
\newblock Tour de {F}rance 2022: Stage analysis.
\newblock {\em Peloton Magazine}.
\newblock 9 June.
  \url{https://www.pelotonmagazine.com/features/tour-de-france-2022-stage-analysis/}.

\bibitem[Young, 1975]{Young1975}
Young, H.~P. (1975).
\newblock Social choice scoring functions.
\newblock {\em SIAM Journal on Applied Mathematics}, 28(4):824--838.

\bibitem[Zhao et~al., 2021]{ZhaoWuChen2021}
Zhao, Y., Wu, F., and Chen, Y. (2021).
\newblock Sensitivity analyses and measurements for group decisions using
  weighted scoring rules.
\newblock {\em International Transactions in Operational Research},
  28(2):959--975.

\end{thebibliography}

\end{document}